\documentclass{optica-article}
\usepackage{comment}

\journal{opticajournal} 

\articletype{Research Article}

\usepackage{lineno}
\usepackage{float}

\begin{document}

\title{Wavelength-Dependent Evolution of Full-Field Transfer Matrices in Photonic Lanterns}

\author{Caleb Dobias,\authormark{1,*} Miguel R{\"o}mer,\authormark{1} Swati Bhargava,\authormark{1} Tara Crowe,\authormark{1} Liza F. Quinn Reyes,\authormark{1} David Smith\authormark{1},  Matias Barzallo,\authormark{1} Daniel Cruz-Delgado,\authormark{1} Sergio Leon-Saval,\authormark{1,2} Stephanos Yerolatsitis,\authormark{1} Miguel A. Bandres,\authormark{1,4} Stephen S. Eikenberry,\authormark{1,3,4} and Rodrigo Amezcua-Correa\authormark{1}}

\address{
\authormark{1}College of Optics and Photonics (CREOL), University of Central Florida, Orlando, FL 32816 USA\\
\authormark{2}Sydney Astrophotonics Instrumentation Laboratory, University of Sydney, Sydney, Australia\\
\authormark{3}Department of Physics | Planetary Sciences, University of Central Florida, Orlando, FL 32816 USA\\
\authormark{4}Florida Space Institute, University of Central Florida, Orlando, FL 32816 USA\\
}

\email{\authormark{*}caleb.dobias@ucf.edu} 

\begin{abstract}
A fiber-based photonic lantern can couple an array of single-mode optical fibers to the guided modes of a multimode fiber, with the mapping between the single-mode fibers and guided modes fully described by a complex-valued transfer matrix. Recent experimental studies have reported strong wavelength-dependent evolution of this matrix in non-mode-selective photonic lanterns, yet a quantitative physical explanation for this behavior has not previously been demonstrated. Here, we present direct measurements of the wavelength-dependent encoding transfer matrix of a photonic lantern across the range \(1525\,\mathrm{nm}\) to \(1575\,\mathrm{nm}\) using off-axis holographic imaging, enabling high-fidelity recovery of both amplitude and phase. Beyond measurement, we introduce a physically grounded propagation model and numerical simulation that quantitatively reproduces the observed wavelength evolution and provides a unified physical explanation for behavior reported in prior experimental work. The model identifies differential modal phase accumulation in the multimode section as the dominant mechanism governing spectral evolution and shows that increasing the length of the multimode end systematically accelerates the phase evolution of the transfer matrix with wavelength. These results establish a direct and predictive link between photonic lantern geometry and spectral response, providing a design framework for tailoring lanterns either to enhance sensitivity to closely spaced wavelengths or to enforce uniform response over broad bandwidths for spectroscopic and imaging applications.
\end{abstract}

\section{Introduction}

\begin{figure}[htbp]
    \centering
    \includegraphics[width=1\linewidth]{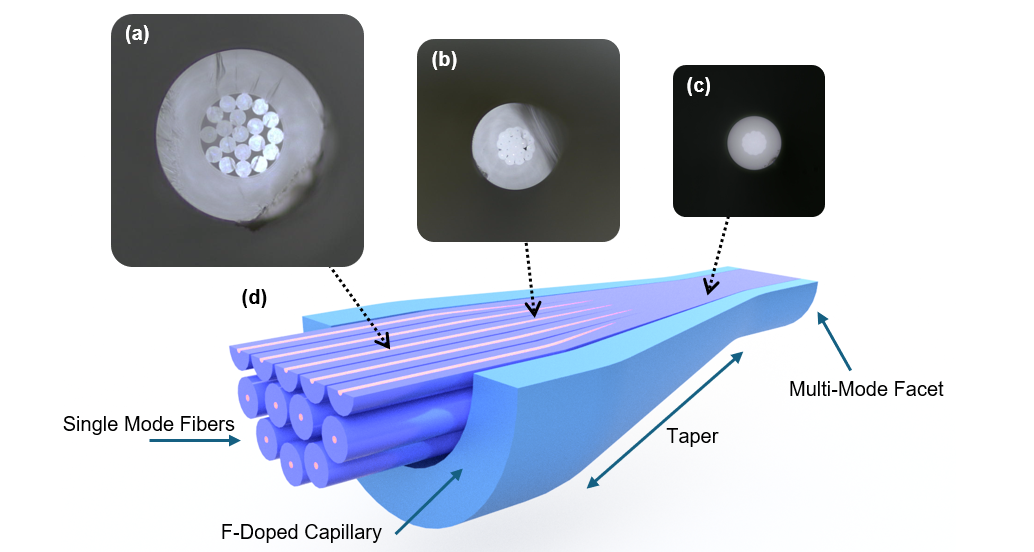}
    \caption{Conceptual layout of an SMF-based PL (not to scale), with images of cross sections of the PL before, during, and after the taper. (a) A cross section of the PL prior to the taper. The single-mode cores are entirely independent and pockets of air exist between the SMFs. (b) A cross section of the PL at the middle of the taper. Here, the SMFs fuse together and interstitial air pockets almost completely collapse. With additional tapering, the SMF cores become smaller and more proximate, where light will couple between the cores as supermodes. (c) The multimode facet of a SMF-based PL. Here, the cores of the SMFs are so small that they no longer guide light or supermode as coupled cores. Instead, the light is contained by the F-Doped capillary, effectively treating the cladding of the SMFs as a fiber core and the capillary as an MMF.  (d) A labeled rendering of a cross section of a SMF-based PL.}
    \label{fig:Lantern Layout}
\end{figure}

Photonic lanterns (PLs) \cite{noordegraaf_efficient_2009, leon-saval_photonic_2010, birks_photonic_2015} are waveguide devices that enable a bidirectional interface between multiple waveguides and a multimode waveguide, most commonly several single-mode fibers (SMFs) and a multimode fiber (MMF). Fig.~\ref{fig:Lantern Layout} illustrates the internal structure of a representative photonic lantern, including the single-mode inputs, the tapered transition region, and the multimode output section. Originally developed in the context of astrophotonics to efficiently collect light in a multimode regime while enabling single-mode filtering \cite{leon-saval_multimode_2005}, photonic lanterns have since found broad application in lidar \cite{ozdur_free-space_2013, ozdur_photonic-lantern-based_2015}, telecommunications \cite{fontaine_geometric_2012, leon-saval_mode-selective_2014}, beam shaping \cite{gross_beam_2019, milne_coherent_2023}, wavefront sensing \cite{norris_all-photonic_2020, lin_experimental_2025}, and sub-diffraction imaging \cite{eikenberry_photonic_2024, kim_-sky_2025}.

A defining feature of photonic lanterns is that they implement a unique (though often effectively random) mapping between the guided modes supported at the multimode facet and the individual single-mode ports. Accurate knowledge of this mapping, or a reliable approximation thereof (for example via neural-network-based inference), is critical for applications such as beam shaping, wavefront detection, and sub-diffraction imaging. Mode-selective photonic lanterns are designed to preserve a one-to-one correspondence between individual single-mode inputs and specific guided modes, leveraging distinct effective indices through dissimilar single-mode cores and preventing mode degeneracy throughout the taper to preserve exclusive coupling \cite{yerolatsitis_adiabatically-tapered_2014,leon-saval_mode-selective_2014, velazquez-benitez_six_2015}. However, fabrication complexity and scalability rapidly become prohibitive as the number of supported modes increases \cite{fontaine_geometric_2012, velazquez-benitez_scaling_2018}. Non-mode-selective photonic lanterns, by contrast, scale readily to much larger mode counts, as there are no such constraints on mode core dissimilarity \cite{noordegraaf_multi-mode_2010, birks_photonic_2012,leon-saval_divide_2017, choudhury_computational_2020}.

A compact and complete description of a photonic lantern is provided by its transfer matrix (TM), a complex-valued matrix that relates fields at the single-mode and multimode ports \cite{fontaine_geometric_2012, velazquez-benitez_scaling_2018, romer_broadband_2025, romer_decoding_2026,taras_illuminating_2026}. Depending on the direction of propagation, this matrix can be interpreted either as an \emph{encoding} matrix, mapping single-mode inputs to multimode output fields (SMF\(\rightarrow\)MMF), or as a \emph{decoding} matrix, mapping multimode inputs to single-mode outputs (MMF\(\rightarrow\)SMF). For non-mode-selective photonic lanterns, both representations exhibit strong wavelength dependence. Consequently, broadband implementation of photonic lantern-based systems requires not only accurate measurement of the relevant transfer matrix, but also an understanding of its evolution as a function of wavelength.

Early efforts to characterize photonic lantern transfer matrices relied primarily on linear approximations or machine-learning-based approaches that recover intensity-only mappings, often over limited wavelength ranges and/or for fewer than the full set of supported input modes \cite{yu_mode-dependent_2016, lin_focal-plane_2022, norris_all-photonic_2020}. More recent work has demonstrated direct measurement of the complex-valued transfer matrix. In particular, a conference proceeding \cite{romer_broadband_2025} and an associated submitted journal article \cite{romer_decoding_2026} measure the \emph{decoding} transfer matrix by injecting controlled multimode fields into the multimode port and recording the resulting single-mode outputs. Complementary to this approach, another manuscript describes the use of off-axis holography to measure the \emph{encoding} transfer matrix of a multicore-fiber-based photonic lantern, directly recovering the complex multimode field produced by excitation of individual single-mode inputs\cite{taras_illuminating_2026}.

In this work, we likewise focus on the \emph{encoding} transfer matrix and measure the full complex-valued mapping from single-mode inputs to multimode output fields for an SMF-based 19-port photonic lantern across a \(50\,\mathrm{nm}\) wavelength range using off-axis holographic imaging. Beyond measurement, we introduce a physically motivated propagation model and numerical simulation that quantitatively reproduces the observed wavelength evolution of the transfer matrix, which has not been previously demonstrated. The model identifies differential modal phase accumulation within the multimode section of the photonic lantern as the dominant mechanism governing spectral evolution and shows that increasing the length of the multimode end systematically enhances the rate of phase evolution with wavelength. This provides a unified physical explanation for the strong wavelength dependence observed both here and in prior experimental studies \cite{romer_broadband_2025,romer_decoding_2026, taras_illuminating_2026}.

Depending on the intended application, different spectral behaviors may be desirable. For broadband systems such as multiplexers \cite{huang_orbital_2022}, minimal wavelength dependence of the transfer matrix is advantageous, whereas in spectrophotometric and wavelength-discriminating applications \cite{kim_potential_2024}, a strong and distinctive wavelength response can be beneficial. The framework presented here establishes a predictive link between photonic lantern geometry and spectral response, enabling the deliberate engineering of photonic lanterns to meet these differing requirements.

\section{Experimental Procedure and Analysis}
\begin{figure}[htbp]
    \centering
    \includegraphics[width=.9\linewidth]{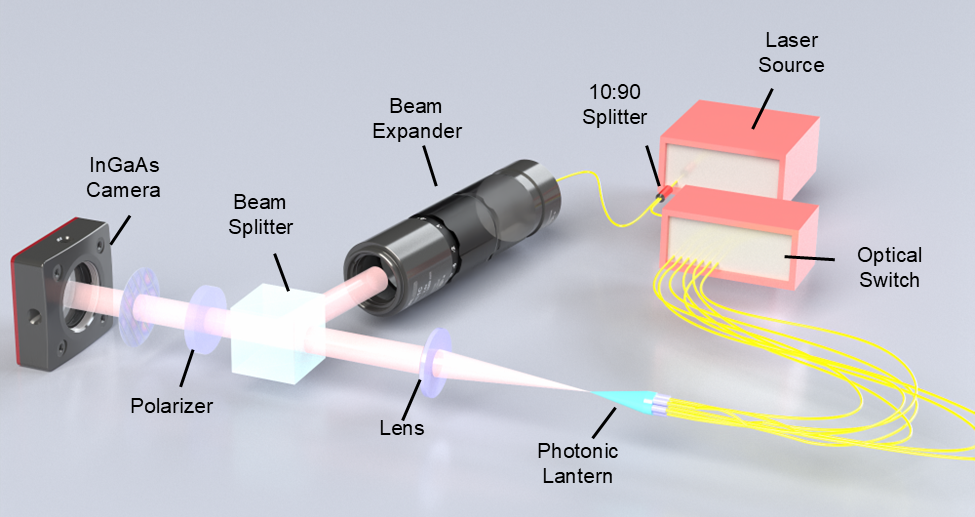}
    \caption{Conceptual layout of the off-axis holography experiment. Light from a tunable laser passes through a 90:10 fiber splitter, such that the reference beam and sample beam have similar relative intensity. Light through the 90\% channel is passed through a beam expander oriented at a $2.3 \deg$ angle with respect to a beam splitter to act as the reference field. Light through the 10\% channel is passed to an optical switch, with the selectable outputs each connected to the individual PL SMF ports. The field that emerges from the PL is imaged by a lens onto the InGaAs camera to act as the sample field. The resulting image on the camera contains a fringe pattern which can be processed to find the phase information of the sample field.}
    \label{fig:Holo Setup}
\end{figure}

The PL studied in this analysis was a 19-core non-mode-selective PL designed for use with $1550\, \mathrm{nm}$ light. This PL was fabricated at the University of Central Florida by packing 19 SMFs in an F-doped silica glass capillary with an OD of $1280\,\mu\mathrm{m}$, an ID of $635\,\mu\mathrm{m}$, and a $\Delta n$ of $0.0058$. The packed capillary was then tapered to a MMF with a core size of $34\,\mu\mathrm{m}$, with a resulting taper length of 5 cm. To capture the phase and intensity of the output fields of the photonic lantern, we use an off-axis digital holography setup in the Mach-Zehnder configuration \cite{van_der_heide_exploiting_2020, lyu_fast_2017} with an optical switch to selectively illuminate the ports of the PL (see Fig.~\ref{fig:Holo Setup}). 

We scanned the tunable laser across the $1525\, \mathrm{nm}$ to $1575\, \mathrm{nm}$ range at $1\, \mathrm{nm}$ intervals, measuring independent holograms for each of the 19 ports at each wavelength separately. As light passes through significant lengths of non-polarization-maintaining fiber, we can consider the light leaving the PL to have a random polarization. To ensure that the polarization of the output of the PL matched the polarization of the reference beam, a polarization filter was placed in front of the PL.  In the case where the polarization of the output of the PL was not aligned with the polarization filter, a motorized fiber polarization controller (not shown) was used to automatically change the polarization of the light entering the optical switch, and thus the PL, until we achieved peak brightness.

\begin{figure}[htbp]
    \centering
\includegraphics[width=1\linewidth]{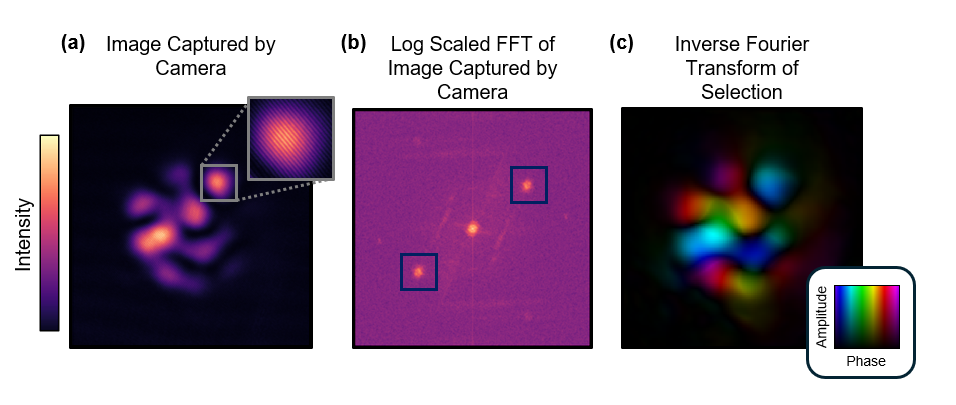}
    \caption{Progression of the hologram image during processing. (a) Real-valued intensity image captured by the InGaAs camera, with a selection expanded to display the interference fringes. (b) The Fourier transform of the real image, log-scaled to showcase the twin images that are highlighted by bounding boxes. Relevant data is retrieved from one of the highlighted twin images using the Fourier Fine-binned method to crop and center the data, with a low-pass (specifically, Butterworth) filter applied to remove high-frequency noise. (c) The electric field recovered by the inverse Fourier Transform of the data retrieved from one of the twin images in (b).}
    \label{fig:Holo Data}
\end{figure}

The process of retrieving the phase information from the images captured in the off-axis holography setup is visualized in Fig.~\ref{fig:Holo Data}. We can write the resulting intensity distribution in the hologram as Eq. (\ref{eq:interference_intensity}). An example distribution is illustrated in Fig.~\ref{fig:Holo Data}(a).

\begin{equation}
    I(r) = |E_R(r)|^2 + |E_S(r)|^2 + E_R^*(r)E_S(r) + E_R(r)E_S^*(r)
    \label{eq:interference_intensity}
\end{equation}

Here, $E_R$ represents the reference field, $E_S$ represents the sample field, and $r$ is the position vector. The first two elements $|E_R |^2+|E_S |^2$ are known as the autocorrelation term or the "DC" term, and do not provide any useful information. However, the last two elements $E_R^* E_S$ and $E_R E_S^*$ (also known as twin images) encode the complex composition of the signal wave. We can describe the reference plane wave as $E_R (r)= Ae^{ikr}$, where $A$ is the amplitude of the plane wave and the wavevector $k$ is proportional to the relative angle between the reference field and the sample field. As the Fourier transform of $E_R$ consists of two delta functions at -k and +k, the Fourier transform of $I(r)$ can then be written as a convolution of the two delta functions and the Fourier transform of $E_S$ as Eq. (\ref{eq:I-DC}).  An example Fourier transform of the received intensity distribution is illustrated in Fig.~\ref{fig:Holo Data}(b).
\begin{equation}
    I(q) = DC(q) + \tilde{E}_S^*(q) \ast \delta(q - k) + \tilde{E}_S(q) \ast \delta(q + k)
    \label{eq:I-DC}
\end{equation}

\begin{figure}[htbp]
    \centering
    \includegraphics[width=1\linewidth]{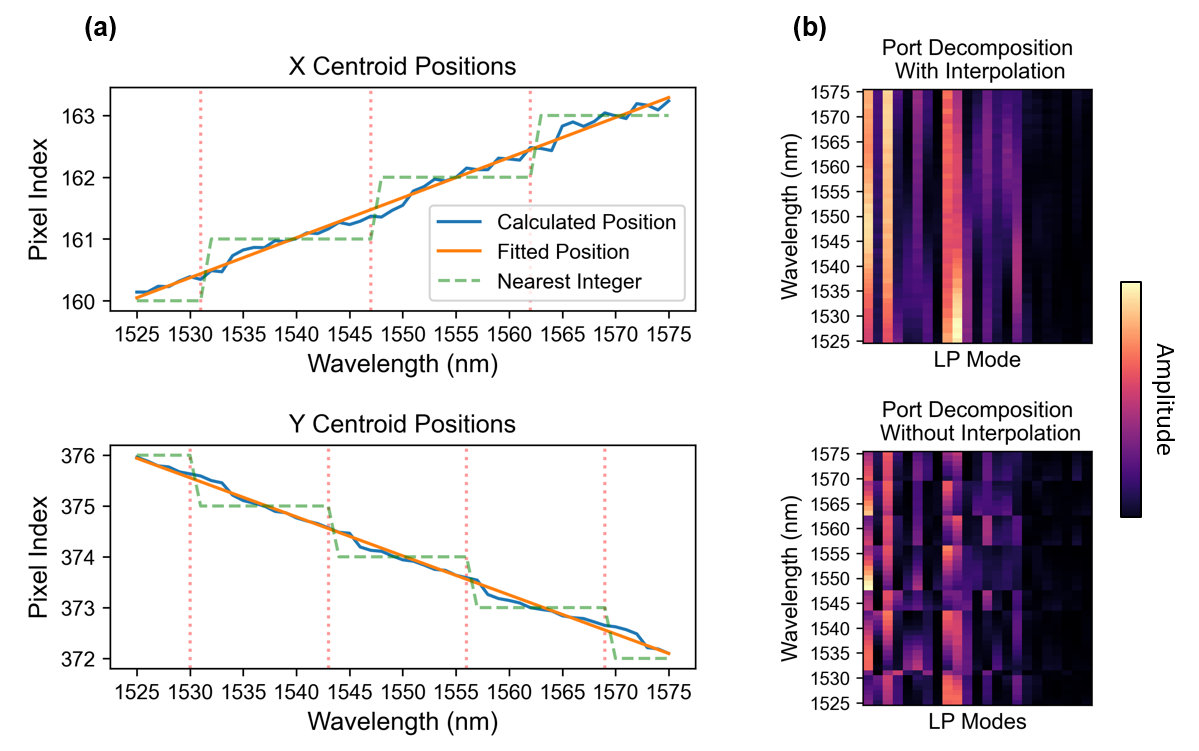}
    \caption{Comparison of a SMF port's LP mode decomposition (see Eq.~\ref{eq:Ck}) with and without using interpolation to select a twin image (see Fig.~\ref{fig:Holo Data}). (a) Calculated X and Y centroid positions of the twin image versus wavelength, along with a linear best-fit function and positions rounded to the nearest integer. Indices where the nearest integer changes are marked with dashed lines. (b) Mode decomposition amplitudes versus wavelength for an SMF port with and without using the Fine-Binning method to interpolate the twin image. Discontinuities in (b) correspond to the changes in centroid positions shown in (a).}
    \label{fig:interpolation}
\end{figure}

 To extract the $E_S$ image and calculate its inverse Fourier transform, we must isolate a subimage centered on a twin. As the wavenumber $k$ changes with wavelength, the offset of $E_S$ also changes. To empirically determine the center offset of the resulting $E_S$, we calculate the image centroid of each $E_S$ for each port and each wavelength by finding the centroid of the convolution of the FFT and a circular mask slightly larger than the size of the relevant data. The resulting convolution should peak at the point where the mask covers the twin image, even if the actual centroid of the data is not at the point of the delta function. We average the centroids for all 19 ports for each wavelength. We then apply a linear best fit to determine an appropriate subpixel estimate for the center of the data. The relatively coarse pixel resolution of the Fast Fourier Transform makes our results subject to artificial variations in the retrieved information due to integer-pixel "jumps" in the image center when the best fit would move the center to an adjacent pixel for a particular wavelength, skewing the recovered information (see Fig.~\ref{fig:interpolation}). To address this, we interpolated the complex-valued $E_S$ using the Fourier Fine-Binning method \cite{ransom_fourier_2002} to center and crop the data on a sub-pixel level without upsampling. Additionally, we also applied a Butterworth filter to remove high spatial frequency noise from the selection. An IFFT of the filtered complex data results in a complex-valued electric field $E_S(r)$ with both phase and amplitude information, illustrated in Fig.~\ref{fig:Holo Data}(c).

The electric field of a wavefront in any multimode waveguide, such as $E_S(r)$, can be described as a linear combination of supported eigenmodes $\Psi_k$ in the waveguide with complex coefficients $C_k$. In this case, we use the retrieved field from the multimode facet of the PL illuminated by a specific single-mode port as $E_S(r)$ and the LP modes expected to be supported by the PL as $\Psi_k$. We calculate $C_k$ via a normalized inner product:

\begin{equation}
    C_k = \frac{\langle E_S(r), \Psi_k(r) \rangle}{\|E_S(r)\| \|\Psi_k(r)\|}
    \label{eq:Ck}
\end{equation}

By calculating the complex coefficient for each LP mode for a retrieved electric field, we obtain the decomposition of the retrieved field in terms of LP modes, or in other words, a row of the TM. To quantify the quality of our decomposition, we generate a recomposition of $E_S(r)$ to compare to the original retrieved field. This recomposition, which we refer to as $E_{\text{Rec}}(r)$, is calculated by summing all the products of all previously calculated LP mode fields by their corresponding complex coefficient:

\begin{equation}
    E_{\text{Rec}}(r) = \sum_{k} C_{k} \Psi_{k}(r)
    \label{eq:E_rec}
\end{equation}

We can consider our recomposition to be a better reconstruction of our retrieved electric field when its fidelity, $\eta$, is higher.  To calculate fidelity, we calculate the squared absolute value of the normalized inner product of the recomposed field, $E_{\text{Rec}}(r)$, and the field we received from our off-axis digital holography, $E_S(r)$:

\begin{equation}
    \eta = \left| \frac{\langle E_{\text{Rec}}(r), E_{\text{S}}(r) \rangle}{\|E_{\text{Rec}}(r)\| \|E_{\text{S}}(r)\|} \right|^2
    \label{eq:eta}
\end{equation}

Several practical effects can affect the recovered field information in the experiment, including the holographic image pixel scale, optical aberrations in the measurement system, and small errors in the estimate of the twin image centers arising from the intrinsic asymmetry of the images. In terms of aberrations, we noted a quadratic phase error of $e^{ik\rho^2}$ where $\rho$ is the radius from the image center. We believe this is a wavelength-dependent feature likely imparted by a slight defocus from the lens imaging the output of the PL onto the camera. To address these effects, several parameters were optimized by iterating across a range of assumed values on a single parameter, selecting the value with the highest fidelity $\eta$ and repeating this process for all parameters, several times, until a global minimum was found for each combination of wavelength and illuminated SMF. The optimized parameters include the position of the retrieved field, the mode field diameter, and the phase factor $k$ in the quadratic phase mask applied to compensate for the quadratic phase error. The optimal parameters were averaged across all SM ports and wavelengths and used for decomposition for data analysis, except for the quadratic phase mask, which instead used a linear fit to address the wavelength dependency of the quadratic phase error.

\section{Results and Discussion}

\subsection{Overview of the Transfer Matrix}

\begin{figure}[htbp]
    \centering
     \includegraphics[width=1\linewidth]{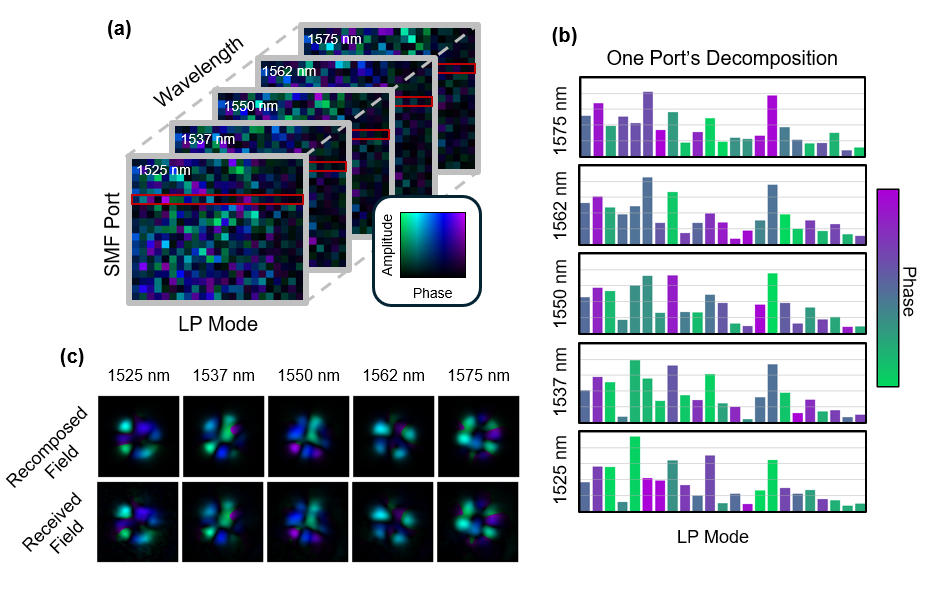}
    \caption{ (a) A visualization of the TM at several wavelengths, with a row of interest outlined in red. Each row, column, and layer corresponds to a particular single-mode port, an LP Mode supported by the multimode portion of the PL, and wavelength, respectively. The relative phase of a matrix unit is depicted by color, while the amplitude is depicted by brightness. (b) A bar graph depicting the modal content of the emphasized port in (a), with amplitude depicted by the height of the bar and phase by the color of the bar. (c) A comparison of the recomposed field, generated from idealized modes calculated from the modal decomposition, and the received field, constructed from off-axis holography.  Here, like in (a), the amplitude and phase are depicted by brightness and color respectively.}
    \label{fig:1550 Transfer Matrix}
\end{figure}

We present visualizations of the resulting transfer matrix (TM), received fields, recomposed fields, and a single port's decomposition at different wavelengths in Fig.~\ref{fig:1550 Transfer Matrix}. Comparing our recomposed fields, generated from idealized LP modes, to those recovered through off-axis holography, we obtain an average fidelity of 98\% with a standard deviation of 0.8\%. Because the absolute phase of each SMF port varies with fiber length, temperature, and mechanical perturbations, we impose a global phase reference on each row such that the LP$_{01}$ mode has zero phase. As expected for non–mode-selective lanterns, the TM exhibits no clear symmetry or modal ordering, consistent with the very small differential propagation constants between modes during tapering (i.e., small $\Delta\beta$). Within the taper itself, the cores become smaller, confining less of the light, and couple with each other as supermodes. Parameters such as the spacing between cores, degree of fusion, fiber geometry, and taper profile can change the effective index of these supermodes as they evolve into LP modes, and, in fact, we expect coupling between supermodes of the same effective index \cite{fontaine_geometric_2012, norris_optimal_2022}. However, fabrication imperfections—such as ellipticity, fiber twisting, surface roughness, or internal stress—can shift modal propagation constants so that $\Delta\beta \approx 0$, leading to strong coupling, as described in classical analyses of tapered lanterns and mode-coupling theory \cite{leon-saval_photonic_2010,ho_mode_2013}, leading to a seemingly random distribution of coupling between a given SM port and LP modes.

Importantly, the qualitative behavior of our measured encoding TMs is consistent with the full-field characterizations recently reported in \cite{romer_broadband_2025}, \cite{romer_decoding_2026}, and \cite{taras_illuminating_2026}. The work in \cite{taras_illuminating_2026} uses multi-wavelength, polarization-resolved digital off-axis holography to show that fabricated lanterns exhibit significant deviations from idealized symmetry, and that the principal modes evolve rapidly with wavelength across tens of nanometers due to subtle geometric and refractive-index variations accumulated during tapering. Likewise, the work in \cite{romer_decoding_2026} demonstrates that the full \emph{decoding} TM of a 19-port lantern decorrelates quickly with wavelength and that the complex-valued mapping exhibits negligible inter-wavelength similarity in many cases. We believe that this decorrelation behavior is not unique to any particular geometry, mode spacing, or tapering as the MCF-based $1550\, \mathrm{nm}$ PL in \cite{taras_illuminating_2026} exhibits the same decorrelation structure as the SMF-based $800\, \mathrm{nm}$ PL in \cite{romer_decoding_2026} and the $1550\, \mathrm{nm}$ SMF-based PL in our work here. However, the previous works do not clarify what effects contribute to this shape, especially pertaining to MM propagation after the taper, a common feature on photonic lanterns \cite{davenport_photonic_2021}.

We further note that although the PL contains 19 single-mode fiber inputs, the multimode facet supports 23 guided LP modes across the wavelength range of interest. This dimensional mismatch implies that the encoding TM obtained here cannot be inverted to yield the decoding matrix, as the transformation is fundamentally non-square and therefore not unitary. Instead, a decoding matrix must be determined independently. This conclusion aligns precisely with the findings in \cite{romer_broadband_2025} and \cite{romer_decoding_2026}, which show that encoding and decoding matrices cannot be derived from one another when $N \neq M$, even in principle, due to intrinsic loss and modal truncation in practical lanterns. Similarly, the multi-wavelength holographic characterization in \cite{taras_illuminating_2026} reveals that real devices often deviate from ideal mode-matching conditions, with additional modes becoming weakly bound or partially guided depending on wavelength and fabrication-induced perturbations.

In this broader context, the behavior of our device is entirely consistent with the current understanding of non–mode-selective lanterns: exact mode-count matching between the SMF array and multimode end is not guaranteed in fabrication, and lanterns are typically operated in the nominally “lossless” direction, where light flows from a larger mode space to a smaller one. Excess modes radiate or couple to cladding states rather than introducing catastrophic cross-talk. The non-square nature of our measured TM therefore reflects expected physical behavior and agrees closely with the structures reported in recent full-field encoding and decoding TM measurements.

\subsection{Wavelength Evolution of the Transfer Matrix}

\begin{figure}[htbp]
    \centering
    \includegraphics[width=1\linewidth]{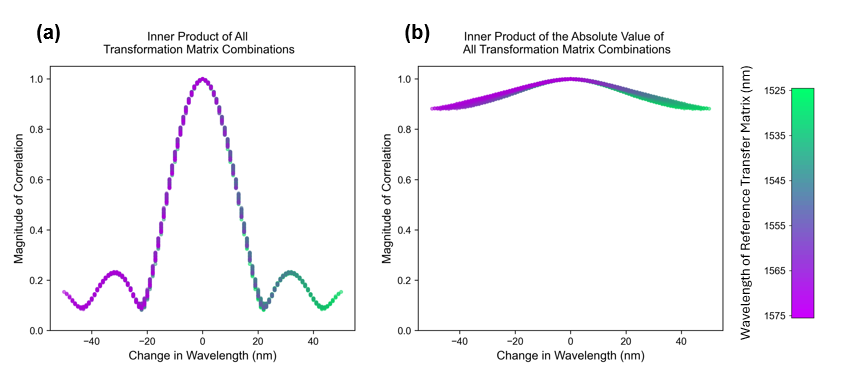}
    \caption{Wavelength evolution of the photonic lantern transfer matrix. (a) Correlation of the full complex-valued transfer matrix as a function of wavelength offset. (b) Correlation of the amplitude-only component of the transfer matrix as a function of wavelength offset. In both panels, the horizontal axis gives the wavelength difference between a reference matrix and a wavelength-shifted test matrix, while the vertical axis shows the magnitude of the corresponding inner product (complex elements in (a); amplitudes only in (b)). Each transfer matrix is used in turn as the reference, with a unique plotting color (shifting from green for 1525 nm to purple for 1575 nm), producing the family of curves shown. The close similarity among these curves indicates that the wavelength evolution is a stable and repeatable property of the lantern, depending primarily on wavelength separation rather than absolute wavelength. The complex-valued correlations decrease much more rapidly with wavelength offset than the amplitude-only correlations, reflecting the fact that the phase content of the transfer matrix varies significantly faster with wavelength than its amplitude distribution.}
    
    \label{fig:Mode Content w Phase}
\end{figure}

The TMs measured at nearby wavelengths exhibit substantial similarity both in phase and amplitude, while those at larger wavelength separations share some similarity in amplitude but change dramatically with phase. To quantify this behavior, we computed the inner product between each TM and all others in the dataset. Plots containing the inner product of each combination of TMs, as well as the inner product of each combination of the absolute values of the TMs, are presented in (Fig.~\ref{fig:Mode Content w Phase}). The resulting correlations initially decrease monotonically with wavelength offset, with a half-power point near 15\,nm separation. Importantly, the correlation curves are essentially independent of the choice of reference wavelength, indicating that the spectral evolution depends primarily on wavelength differences rather than absolute wavelength.

After the correlation reaches a minimum near zero, it increases again at larger separations of approximately 35\,nm, suggesting a partial recurrence of modal phase relationships. This behavior is consistent with differential phase accumulation among the guided modes in the multimode section: as the relative phases diverge, the overlap between TMs decreases until subsets of modes re-align in phase, producing secondary maxima. When the absolute values of the TM elements are used instead of the complex values, the correlations remain significantly higher (around 88\% over 9\,nm). While we cannot assume anything about phase propagation during the taper itself, it is safe to assume that the effect of MM propagation is phase only, as the dominant influences on mode mixing in MMF are bending and twisting \cite{palmieri_mode_2025}. As bending and twisting are prevented by the MM portion of the lantern being packaged in an epoxy-filled ferule, we believe that there is some wavelength-dependent intensity decorrelation happening during the taper itself.

\begin{figure}[htbp]
    \centering
    \includegraphics[width=1\linewidth]{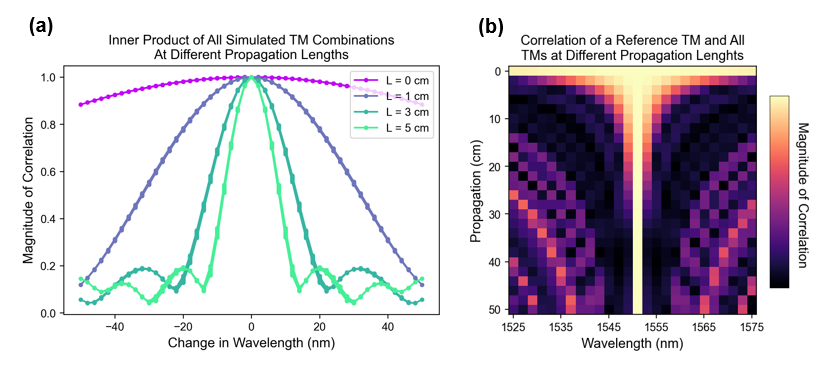}
\caption{Various plots showcasing the behavior of the TMs of a simulated photonic lantern. (a) Overlay of the absolute correlation between all TM combinations to compare with Fig. \ref{fig:Mode Content w Phase}(a).  Different propagation lengths are shown in different colors. The curves coincide closely, further supporting the idea that correlation depends mainly on $\Delta\lambda$ rather than absolute wavelength. (b) Absolute correlation between the TM at $1551\, \mathrm{nm}$ and a TM at the wavelength given by the x-axis and the length between the taper and multimode facet.}
    \label{fig:TM_Overlap_Simulation}
\end{figure}

To further investigate the origin of this spectral behavior, we simulated a 19-port photonic lantern using the Taper Reference Frame method \cite{tschernig_efficient_2026} and calculated TMs at the same wavelengths as the experiment. The idealized propagation in the MM section of the PL introduces only a phase shift depending on the relative effective index of the modes, so we also compute the propagation constants of each relative mode and wavelength. The new TM after MM propagation is expressed in Eq. (\ref{TM_L}).
\begin{equation}
    \mathbf{TM}_L(\lambda) = \mathbf{TM}_0(\lambda)\, e^{j \boldsymbol{\beta}(\lambda) L}
    \label{TM_L}
\end{equation}
Here, $\mathbf{TM}_L(\lambda)$ denotes the TM after propagation over a distance $L$ in the MMF, 
and $\mathbf{TM}_0(\lambda)$ is the TM at $L=0$. $\mathbf{TM}_0(\lambda)$ is a 
$19 \times 23$ matrix, corresponding to the 19 input SM cores and 23 LP modes supported by the multimode end. The term $e^{j \boldsymbol{\beta}(\lambda) L}$ is a $23 \times 23$ diagonal matrix containing the phase factors corresponding to the propagation constants, $\boldsymbol{\beta}(\lambda)$, of the 23 eigenmodes at wavelength $\lambda$.
To quantify the correlation of the TMs at different wavelengths, we again calculate the inner product between all possible TM pairs at different values of $L$.  Plots containing the inner product of each combination of these TMs at different lengths, as well as a visualization of the magnitude of correlation between the TM at 1551 nm and all wavelengths, are presented in Fig.~\ref{fig:TM_Overlap_Simulation}. When $L = 0$, the correlation between TMs at different wavelengths evolves very slowly, due to mode propagation changing during the taper itself. As $L$ increases, we notice that the evolution of the TM increases rapidly, with the main contributing factor being separation between two wavelengths.  We attribute this property to the difference in the effective index between LP modes being significantly wavelength-dependent, such that even relatively short MM propagation causes significant changes to the evolution of the TM.

The correlation again follows a sinc-like dependence centered at the reference wavelength, with the highest correlation occurring at zero wavelength offset ($\Delta\lambda = 0$). However, at sufficient distances, the damping and peak spacing are not consistent, and periods of correlation will spike or change spacing (most noticeable at L = 5 cm). We again observe an important phenomenon that the correlation behavior is primarily governed by the wavelength difference ($\Delta\lambda$) rather than the absolute wavelength, as indicated by the strong overlap of all curves. 

These results collectively indicate that modal propagation in the multimode section sets the fundamental spectral behavior of the photonic lantern, while taper-region imperfections contribute secondary effects. Although wavelength-dependent changes in amplitude are present, they are comparatively small and occur over longer spectral scales. The dominant wavelength dependence arises from intermodal phase evolution, which can be directly controlled through the length and geometry of the multimode region, which is to say, the FWHM of the sinc-like profile decreases with increasing multimode length. This provides a practical means to engineer lanterns for either broadband stability or strong wavelength discrimination, depending on the needs of the intended application.

Of course, there will always be some wavelength dependencies that are intrinsic to the taper. Our simulation shows that with no propagation, the PL's TM drops to 88\% correlation over a wavelength difference of $50\, \mathrm{nm}$. However, this is due to phase only decorrelation during the taper. If we compare the absolute values of the TMs, we see a correlation near unity (99.92\%) over the same wavelength difference. We can expect, therefore, that longer tapers will likely contribute to a faster evolution of the TM with wavelength. However, unlike the simulated data, the correlation of the absolute values of the TMs we retrieved from the physical data drops to 88\% across $50\, \mathrm{nm}$, indicating wavelength-dependent mode scrambling within the taper not seen in our simulation. Despite this decorrelation, the dominant contributor to the evolution of the TM is still MM propagation. As PLs, regardless of manufacturing, include a transition from several waveguides to a MM waveguide, we expect there to be some unavoidable wavelength dependency from the transition itself, especially close to the MM waveguide. We can, however, control MM propagation by cleaving or polishing down the PL immediately after or with intentional distance from the taper.

The extent to which the wavelength-dependent mode mixing within the taper can be affected by the taper length or exaggerated by disorder such as the twisting of fibers within the taper is left for future work. Also left for future work is the physical verification of this process through the manufacturing of a PL with a significantly long multimode section that is cut back and characterized at several lengths. While the deliberate use or avoidance of MM propagation in PL structures provides new manufacturing challenges, we believe considering the effect of multimode propagation is worthwhile in the manufacturing of PLs for broadband applications.

 \section{Conclusion}
 We have measured the full complex-valued, wavelength-dependent encoding transfer matrix of a 19-port single-mode fiber photonic lantern between \(1525\,\mathrm{nm}\) and \(1575\,\mathrm{nm}\) using off-axis digital holography. The measured matrices exhibit rapid spectral evolution in phase and relatively slow variation in amplitude, consistent with the behavior reported in recent full-field characterizations of photonic lanterns, including the multi-wavelength encoding measurements in \cite{taras_illuminating_2026} and the decoding-matrix results in \cite{romer_broadband_2025} and \cite{romer_decoding_2026}. To offer an explanation of this behavior not previously investigated, we developed a propagation model that quantitatively reproduces the observed spectral evolution. The model identifies differential modal phase accumulation in the multimode section as the dominant mechanism governing the wavelength dependence, while wavelength-dependent amplitude variations likely arise mainly from secondary fabrication-induced effects. The simulated transfer matrices show the same decorrelation behavior as the measured matrices, including the dependence on wavelength separation and the presence of secondary correlation maxima, indicating that multimode-region propagation provides a physically-grounded, predictive explanation for the structure of the measured matrices. These results establish a direct connection between the geometry of the multimode region and the spectral response of non–mode-selective photonic lanterns. Because the rate of phase evolution increases with multimode-section length, the spectral characteristics of a lantern can be intentionally tailored by adjusting this length. Devices requiring stable broadband performance may benefit from minimizing the multimode region, while applications that rely on strong or distinctive wavelength signatures may exploit longer multimode sections. This understanding provides a foundation for designing photonic lanterns with controlled and application-specific spectral behavior.

\begin{backmatter}

\bmsection{Funding}
This material is based upon work supported by the Air Force Office of Scientific Research under award number FA9550-24-1-0332. 

\bmsection{Acknowledgment}
The authors would like to acknowledge Timothy Bate and Mark Whitledge for their insight and support during the data processing and the writing of this paper as well as Christopher Betters for his insight into the fabrication of photonic lanterns. 
The authors also acknowledge support from the University of Central Florida SPICE Academic Excellence Program and the PHAST Jumpstart Initiative.

\bmsection{Disclosures}
Sergio Leon-Saval is a co-founder of Sydney Photonics, a company specializing in the production of photonic lanterns. No company resources were used in the fabrication or analysis of the photonic lantern detailed in this manuscript, nor did this affiliation influence the reporting of this device.

\bmsection{Data availability} Data underlying the results presented in this paper are not publicly available at this time but may be obtained from the authors upon reasonable request.

\end{backmatter}

\bibliography{references}
\end{document}